# The rotational memory effect of a multimode fiber

**L. V. Amitonova, A. P. Mosk and P. W. H. Pinkse**

*Complex Photonic Systems (COPS), MESA+ Institute for Nanotechnology, University of Twente, PO Box 217, 7500 AE Enschede, The Netherlands*
*l.amitonova@utwente.nl*

**Abstract:** We demonstrate the rotational memory effect in a multimode fiber. Rotating the incident wavefront around the fiber core axis leads to a rotation of the resulting pattern of the fiber output without significant changes in the resulting speckle pattern. The rotational memory effect can be exploited for non-invasive imaging or ultrafast high-resolution scanning through a multimode fiber. Our experiments demonstrate this effect over a full range of angles in two experimental configurations.

## 1. Introduction

Fiber-optic technologies provide a broad variety of powerful tools for biological and medical applications [1,2] as well as for remote sensing, communication [3–5] and recently, as a platform for quantum information processing [6]. Optical imaging via narrow probes and fiber bundles allows high-speed fluorescence microscopy in freely moving mice [7], long-term deep-tissue measurements of neural activity [8], *in vivo* monitoring of chemically specific markers [9] and optical control of neural circuits [10]. However, the spatial resolution of widely used fiber bundle probes is limited by the diameter of the individual fibers and by the distance between them. The pixelization in practice leads to a resolution of only 2.5-3 μm [11]. GRIN lenses in miniaturized microscopes for life-science applications also provide a resolution of only about 2.5 μm [12]. Multimode (MM) fibers potentially offer much better resolution and a minimal cross section [13]. However, unavoidable coupling of the many modes of a MM fiber prevents direct image transmission. For this reason several groups have turned to wavefront shaping, in which light is focused through a scattering medium using spatial light modulators (SLMs) [14–16]. It was shown that interference of different fiber modes on the output of a standard multimode fiber can also be controlled by wavefront shaping [17,18]. Multimode-fiber-based microendoscopy could reach the fundamental minimum probe dimensions with diffraction-limited spatial resolution without pixelization and is suitable for three-dimensional imaging of fluorescent objects since it is possible to scan in the longitudinal direction.

To compensate for changes in mode coupling because of bending, temperature or strain changes, fast and constant reoptimization is needed. An iterative implementation with a fast (32 kHz) amplitude digital micromirror array (DMD) instead of a phase SLM has demonstrated re-focusing through a multimode fiber in 37 ms [19], implying more than two hours for a 512x512 pixel image. Another disadvantage of current wavefront-shaping fiber imaging methods is that they are invasive because they all (wavefront shaping, transmission matrix measurements, phase conjugation) require a detector behind the fiber [20–22].

A non-invasive procedure of imaging through opaque scattering layers was demonstrated by Bertolotti *et al*. [23]. This method exploits the angular memory effect [24] to obtain a detailed image through opaque scattering media. The angular memory effect in scattering media means that tilting the incident beam over small angles $\theta$ doesn't change the resulting speckle pattern, but only translates it over a distance $\Delta r \approx \theta d$. This memory effect has emerged as a powerful tool for the non-invasive reconstruction of images from scattered light [23,25–30]. Formally, the angular memory effect is a consequence of the $C_1$ correlation coefficient [31] and is significant in a thin and wide scattering layer. The geometry of a multimode fiber is different: long and narrow; the angular memory effect is absent.

Here, we show the *rotational* memory effect in a multimode fiber. Rotating the incident wavefront around the axis of the fiber core leads to a rotation of the resulting pattern of the fiber output, while maintaining a high degree of correlation in the resulting speckle pattern. Our experiments demonstrate this effect over a full range of angles, 360°.

## 2. Experimental setup

Our experiments are performed on a multimode fiber [Thorlabs, FG050UGA] with silica core of 50 μm diameter, NA = 0.22 and a length of 12 cm. Such a fiber sustains approximately 1500 modes. We use the continuous-wave linear polarized output of a He-Ne laser with a wavelength of 633 nm (Fig. 1). A single-mode fiber is used to clean the laser mode and to expand the laser beam to match the surface of our spatial light modulator. To control phase and amplitude on the input facet of the fiber with high speed we use a 1920x1200 Vialux V4100 DMD. Each mirror of the DMD can be set to two different tilt angles, therefore creating a binary amplitude mask. The Lee amplitude holography method [32] is used to create the desired wavefront; the more powerful superpixel method would have worked as well [33]. Lenses L1 and L2 are placed in a 4f-configuration to image the phase mask on the back focal plane of a 20x (NA = 0.4) objective that couples the light into the MM fiber. A pinhole in the Fourier plane blocks all the diffraction orders except the 1st, encoding the desired phase distribution. An objective 40x (NA = 0.65) is used to collect light from the fiber output. Two cameras image both the input and the output facets of the MM fiber in a polarization-insensitive way.

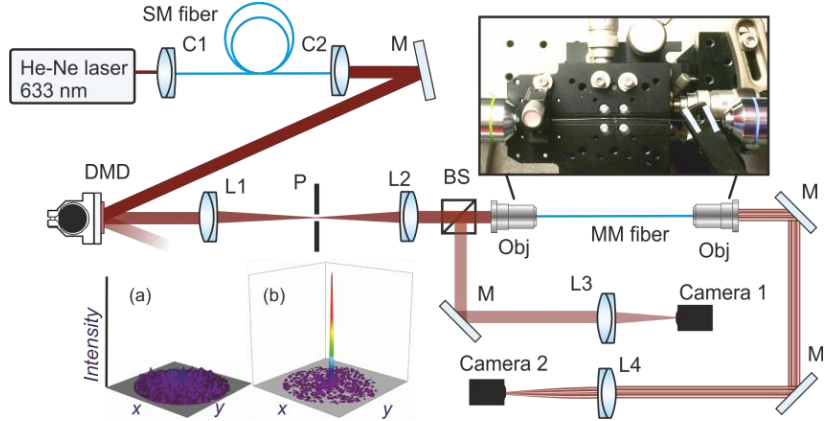

Fig. 1. Experimental setup to demonstrate the rotational memory effect in a multimode fiber (MM, multimode fiber; DMD, digital micromirror device; SM, single mode fiber; M, mirrors; L, lenses; BS, beam splitter; Obj, objectives). Intensity pattern behind the MM are shown for (a) a random input (speckle) and (b) the output after optimization of the input wavefront for a single focus. The picture shows the MM fiber (thin horizontal line) in the setup.

A typical multimode fiber output looks like a random speckle pattern, as can be seen in Fig. 1(a). The speckle is caused by the large number of contributing modes which are coupled in a very complex and therefore seemingly random way, already for a perfect fiber. By wavefront shaping we can produce a diffraction limited focal spot on the output facet (Fig. 1(b)).

## 3. Results and discussion

In the first set of experiments we examine the speckle correlation during rotation of the focal spot on the fiber input. First, we determine the fiber core axis and parameters of the input beam by imaging the fiber input with Camera 1. The FWHM of the focused beam is 1.1 μm. By writing gratings with different *k*-vectors over the DMD area, we adjust the position of the focused laser beam on the fiber input facet. The experimental procedure is the following. First, we place the focal spot at a distance $r_m$ = 13 μm from the fiber core axis (core radius $r_0$

= 25 μm) with angle $α = 0°$ with the vertical (inset in Fig. 2(a)) and measure the speckle pattern output with Camera 2. Then we change the angle $α$, measure the new speckle pattern, rotate it back to the same angle and find the correlation coefficient between the two speckle images. Results of such an operation performed for angles $α = -180 … 180°$ are presented in Fig. 2(b) and in Media 1 [34]. The blue dots in Fig. 2(a) represent measurements of the correlation coefficient $r(s_0,s_α)$ between the speckle pattern obtained at the input position at $α = 0°$ and the speckle patterns corresponding to angles $α$ varying from -180 to 180° with a step of $Δα = 2°$ in accordance to the procedure described above. The correlation coefficient is normalized in such a way that 1 corresponds to identical images $r(s_0,s_0)$ and 0 corresponds to the average correlation coefficient $r(s_i,s_j)$ between two random speckle patterns at the fiber output: $r(s_0,s_α) = [r_{exp}(s_0,s_α) − r_{exp}(s_i,s_j)]/\max[r_{exp}(s_0,s_α) − r_{exp}(s_i,s_j)]$, where the subscript exp refers to values obtained from experimental data. As can be seen in Fig. 2(a), the correlation coefficient never drops down to that of a completely uncorrelated speckle patterns and has a minimum value of about 0.3 around ±90°. This means that we experimentally observe the rotational memory effect in a multimode fiber over a full range of angles (360°). The revival of the cross correlation around ±180° indicates that the deviations from the perfect cylindrical symmetry do not completely break the mirror symmetry of the fiber. Defining an angular coefficient $Δα_{0.5}$ as the full width at which the correlation coefficient drops to a level of 0.5, we find $Δα_{0.5} = 100°$ in our experiments.

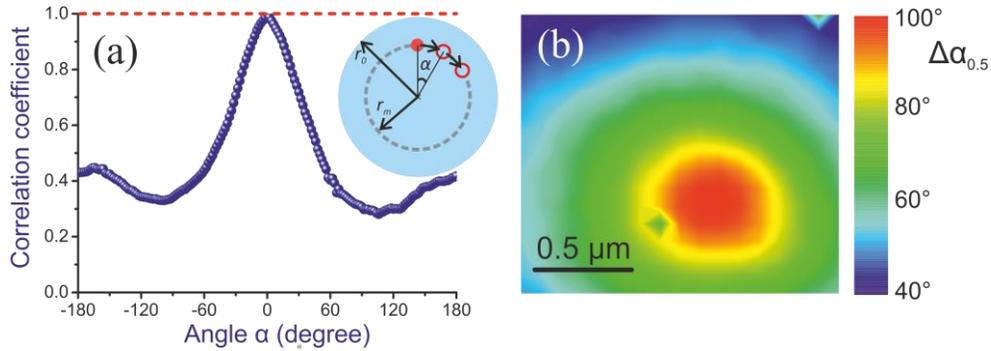

Fig. 2. (a) The correlation coefficient of the speckle patterns as a function of the angular position of the input spot, α. Blue dots – experimental data, red dashed line – theoretical calculation for an ideal fiber. The inset shows a sketch of the scanning procedure. The filled red circle represents the initial position of the focused beam on the fiber input. The open red circles two consecutive positions with an angular step $Δα = 2°$. The radius of the fiber core is $r_0 = 25$ μm, the scan radius $r_m = 13$ μm. The fiber output during this rotation is shown in Media 1 [34]. The full width at which the correlation coefficient drops to a level of 0.5 amounts to $Δα_{0.5} = 100°$ in our experiment. (b) Map showing the experimentally observed $Δα_{0.5}$ as a function of the position of the rotation axis.

A theoretical calculation of the mode propagation has been performed to prove the existence of the rotational memory effect in a MM fiber and to find a theoretical limit of the visibility of this effect. We have found propagation constants $β$ for every mode of the MM fiber by numerical solving the characteristic equation for a dielectric step-index cylindrical waveguide. In addition we have analytically calculated the spatial profiles of every eigenmode and their resulting interference pattern on the fiber output as a function of the position of the focused input Gaussian spot [35]. Using these theoretical 'speckle patterns' we plotted the correlation coefficient as a function of the angular position $α$ of the input spot in accordance to the experimental procedure (red line in Fig. 2(b)). The calculation verifies the existence of a perfect rotational memory effect in an ideal MM fiber. Clearly, in case of a perfect cylindrical symmetry, angular momentum is conserved and rotating the input pattern rotates the output pattern over the same angle but further keeps it unchanged. Differences between theory and experimental results can be explained by small imperfections in the fiber core geometry, small amounts of fiber bending or scattering from inhomogeneities [36].

The angular extent of the memory effect in the MM fiber critically depends on an accurate determination of the center position used for the rotation procedure. It should coincide with the true fiber core axis. To show this, measurements of the correlation coefficient as a function of the angular position α are performed for different rotation centers in the area 1.6 × 1.6 µm near the fiber core axis. The fiber core axis can be determined with an accuracy of about 1 µm from images from Camera 1 (see Fig. 1). The results are shown in Fig. 2(b) where the FWHM of the correlation coefficient graph $\Delta\alpha_0$ is plotted for different rotation center position scanned with a step size of 0.2 µm. When the position of the rotation coincides with the position of fiber core axis, $\Delta\alpha_0$ reaches its largest value of 100°. An error of only about 500 nm decreases the visibility of the cross correlation significantly (Fig. 2(b)). Vice versa, this method allows to determine the axis of the fiber core with an accuracy of better than 100 nm.

In a second set of experiments we apply the rotational memory effect for fast scanning of a tightly focused spot at the MM fiber output. To create a focus spot on the fiber output, we use an optimization algorithm [14]. The area of the DMD is divided into 36×36 elements of 15×15 pixels each. The phase of every segment is adjusted independently by changing the phase of the grating in that segment. We determine the optimal phase for a single segment by changing the phase and measuring the intensity in the desired region with three equidistant points (0, $2\pi/3$ and $4\pi/3$), and then interpolate using a cosine function. This procedure creates a tightly focused spot at the desired position on the fiber output facet as shown in Fig. 1(b). We create a diffraction-limited spot at the output facet of the fiber with a FWHM 1.4 µm at a position of $r_m = 13$ µm and α = 0° in the coordinate system presented in the inset of Fig. 2(a).

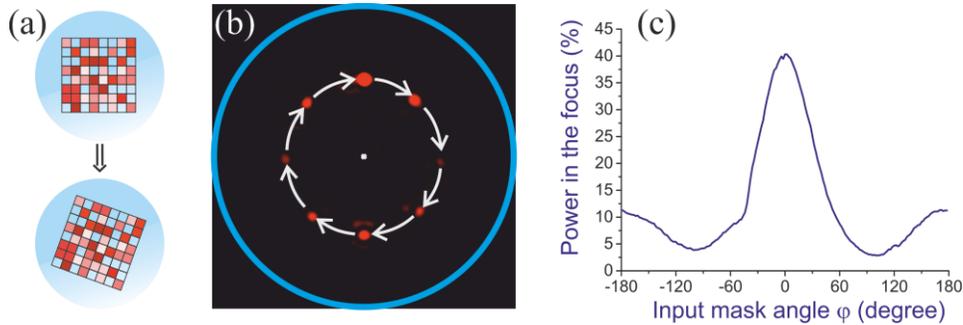

Fig. 3. (a) Method for fast scanning of a tightly focused spot at the fiber output by using the rotational memory effect in a MM fiber. The blue disc represents the ⌀50 µm fiber core; the red square pattern represents an optimal phase mask that focuses a laser beam at the desired point on the fiber output. Rotation of the phase mask around the fiber axis leads to rotation of the focus spot at the fiber output. (b) Sum of eight pictures taken from the output fiber facet during the rotation of the optimized input phase pattern. Arrows represent the rotation direction; the blue circle represents the edge of the fiber core (see Media 2 [37]). (c) Power in the focal spot on the fiber output during the scanning procedure as a function of the angle of the optimized phase pattern on the fiber input.

So far, modification of the position of the focal spot on the fiber output requires repeating the whole optimization procedure, which takes a significant amount of time. By exploiting the rotational memory effect we need only a single optimal phase pattern to scan the focal spot over a wide range of angles. To demonstrate this, we rotate the optimized phase pattern around the fiber core by an angle φ as shown in Fig. 3(a) using a bilinear interpolation procedure. The blue disc represents the fiber core and the red square pattern represents an optimal phase mask. The fiber output during this rotation is shown in Media 2 [37]. The sum of the eight pictures taken from the fiber output is shown in Fig. 3(b). Arrows represent the rotation direction; the blue circle represents the edge of the fiber core. Despite the drop in intensity around φ = ±90°, the focal spot is always significantly brighter than the background. This clearly demonstrates the rotation memory effect over a range of angles from -180 to +180°. Consequently, we require only a single optimized phase pattern to scan the focal spot over a full circle on the output facet of MM fiber.

To characterize the focal spot at the fiber output during rotation, we perform measurements of the size of the focal spot and its intensity. First, we estimate the focal size by measuring the FWHM of the focus spot at the fiber output as a function of the input fiber mask angle φ. The angle is changed from -180 to +180° with a step of 2°. We find that the focus size remains almost constant over a full circle with a FWHM = 1.40 μm and a standard deviation of 0.08 μm. Second, we characterize the power in the focal spot. We define the power in the focus as the light power in the round area centered on the focal spot maximum and a radius equal to the FWHM of the focused beam divided by the total power transmitted through the fiber. Fig. 3(c) shows the power in the focus as a function of input fiber mask angle φ, which is scanned from -180 to +180° with a step size of 2°. As can be seen from this graph, the power in the focus varies from 4% to 40% during rotation. Despite these changes in focus intensity, a significant fraction of the light remains in the focal point during the whole rotation and the focus is at all times the brightest spot in the output plane. The revival of the power around ±180° corresponds to the results obtained in the first set of measurements (Fig. 2(a)) and indicates that the deviations from the perfect cylindrical symmetry do not completely break the mirror symmetry of the fiber.

The wavefront-shaping procedure used for the creation of a focal spot on the fiber output takes about 3 min in our experiments. The limiting factor is the speed of the camera with an acquisition frame rate of about 20 fps. However, the DMD allows ultrafast wavefront shaping and requires only 44 μs to change the phase mask. It was shown [19] that with a DMD, only 37 ms is required to find the optimized phase pattern for one focal point. Consequently, the standard scanning procedure over a fiber core diameter of 50 μm with a step size of 1 μm would require approximately 2000 points to optimize, taking 74 s. By exploiting the rotational memory effect and the scanning procedure presented above, it will take only about 1 s to get an image with the same spatial resolution. Hence, the scanning procedure based on the memory effect is almost two orders of magnitude faster, opening up new horizons for real-time imaging through a multimode fiber.

## 4. Conclusions

We have demonstrated the rotational memory effect in a multimode fiber. Rotating the incident wavefront around the fiber core axis leads to a rotation of the resulting output pattern over a full range of angles without any significant changes. Exploiting the memory effect opens new ways to non-invasive imaging through a multimode fiber by using speckle correlations as well as ultrafast scanning of the focused laser spot on the output fiber facet. It might also allow transmission of high-dimensional quantum states via multimode fibers for Quantum-Secure Authentication with remote readout [38].

**Acknowledgements**

We thank Adrien Descloux, Ravitej Uppu, Cornelis Harteveld and Willem Vos for support and discussions. This work has been funded by FOM and NWO (Vici grant), ERC and STW.